\documentclass[pdflatex,sn-mathphys]{sn-jnl}

\jyear{2022}%
\usepackage{siunitx}
\usepackage{multirow}
\usepackage{hyperref}
\usepackage{color}
\usepackage{siunitx}
\theoremstyle{thmstyleone}%
\theoremstyle{thmstyletwo}%

\theoremstyle{thmstylethree}%

\begin{document}

\title[Measuring the Cosmic X-ray Background accurately]{Measuring the Cosmic X-ray Background accurately}


\author*[1]{\fnm{Hancheng} \sur{Li}}\email{hancheng.li@unige.ch}
\author*[1]{\fnm{Roland} \sur{Walter}}\email{roland.walter@unige.ch}
\author*[1]{\fnm{Nicolas} \sur{Produit}}\email{nicolas.produit@unige.ch}

\author[2]{\fnm{Fiona} \sur{Hubert}}

\affil*[1]{\orgdiv{Department of Astronomy}, \orgname{University of Geneva}, \orgaddress{\street{16 Chemin d'Ecogia}, \city{Versoix}, \postcode{CH-1290}, \country{Switzerland}}}

\affil[2]{\orgdiv{EPF-Ecole d'ingénieur-e-s}, \orgname{}, \orgaddress{\street{55 Av. du Président Wilson}, \city{Cachan}, \postcode{FR-94230}, \country{France}}}

\abstract{Synthesis models of the diffuse Cosmic X-ray Background (CXB) suggest that it can be resolved into discrete sources, primarily Active Galactic Nuclei (AGNs). Measuring the CXB accurately offers a unique probe to study the AGN population in the nearby Universe. Current hard X-ray instruments suffer from the time-dependent background and cross-calibration issues. As a result, their measurements of the CXB normalization have an uncertainty of the order of $\sim$15\%. In this paper, we present the concept and simulated performances of a CXB detector, which could be operated on different platforms. With a 16-U CubeSat mission running for more than two years in space, such a detector could measure the CXB normalization with $\sim$1\% uncertainty.}




\maketitle

\newpage

\section{Introduction}\label{sec:intro}
	
The diffuse Cosmic X-ray Background (CXB) was discovered during a rocket flight \citep{1962PhRvL...9..439G} together with Sco X-1. A Moon observation with \textit{ROSAT} \citep{1991Natur.349..583S} showed the bright side of the Moon reflecting Solar X-rays whereas the dark side revealed a shadow of the CXB, demonstrating its extrasolar origin \citep[although the solar wind induced charge exchange emission from the Heliosphere could contribute in soft X-rays,][]{2009ApJ...691..372S}. The high isotropy of the CXB, measured by \textit{Uhuru}, confirmed an extragalactic origin \citep{2001ApJ...551..624G}. 
	
Thanks to the focusing capabilities of soft X-rays instruments and the help of deep field surveys by \textit{XMM-Newton} \citep{2001A&A...365L..45H} and \textit{Chandra} \citep{2002ApJS..139..369G, 2003AJ....126..539A}, up to 93\% of the extragalactic CXB below \SI{10}{keV} has been resolved into point-like Active Galactic Nuclei (AGNs) \citep{2003ApJ...588..696M, 2005MNRAS.357.1281W}. The sensitivity of the current hard X-ray instruments is not enough to resolve the CXB at hard X-rays where the bulk of its emission lies. As a result only bright AGN could be detected and the fraction of CXB resolved remains less than 39\% \citep{2016ApJ...831..185H} at $\sim$\SI{30}{keV}. 
	
An additional Galactic component is observed below $\sim$\SI{2}{keV} \citep{1969Natur.224..134C} with a extension of $\sim\SI{10}{^{\circ}}$ around the Galactic plane \citep{1985Natur.317..218W} and explained with the emission of faint cataclysmic variables \citep{2009Natur.458.1142R}, thermal emission from ionized gas in the local bubble beyond the neutral interstellar medium \citep{2017ApJ...834...33L}, and scattering of X-rays on illuminated diffuse gas clouds \citep{2014A&A...564A.107M}.
	
X-ray instruments suffer from the time-dependent background and sensitivity change caused by solar activities (soft X-rays), cosmic particles, instrument aging and inaccurate in-orbit calibration resulting in systematic uncertainties. As a result, the CXB intensity measured by many experiments disagrees with each other by up to 10\%-15\% even though its spectral shape is rather well constrained.
	
Extrapolating stacked AGN spectra accounting for the CXB spectrum below \SI{10}{keV} does not reproduce the CXB at hard X-rays. A large population of Compton thick sources (where the density of obscuring material is high enough - N$_H > 10^{24}$ cm$^{-2}$ - for Compton scattering to dominate) has been hypothesized to fill the gap \citep{1989A&A...224L..21S, 1995A&A...296....1C}, however, that population was not detected in deep X-ray surveys \citep{2015A&A...573A.137L, 2015ApJ...808..184M, 2015ApJ...808..185C, 2018MNRAS.480.2578L} nor at hard X-rays \citep{2009MNRAS.399..944M, 2016A&A...594A..73A}.
	
The comparison of the observed CXB with the results of synthesis models puts constraints on the fraction of AGNs with different degrees of obscuration and reflection \citep{2016A&A...590A..49E}. The relation between reflection and absorption contains important information on the average AGN inner geometry. The systematic error in the CXB normalization is however a major source of uncertainty in these models.
	
To measure the CXB accurately, instrumental background modeling, energy and detection efficiency calibration are meticulously required. In this context, the MVN (Monitor Vsego Neba) instrument was proposed by the Space Research Institute of the Russian Academy of Sciences \citep{2021ExA...tmp...40S}. A cylindrical multi-layer collimator protects an inner spectrometer from receiving off-axis photons up to a certain energy threshold, and a rotating obturator periodically shields the aperture of the collimator to modulate the Field of View (FoV) to discriminate the CXB flux from other components and backgrounds.
	
In this work, we present an improved detector concept which mainly consists of an array of collimated spectrometers with rotating obturators on top of the apertures. The detector could be operated on a space station, on a small satellite or even on a CubeSat. The science goals of this detector are discussed in Sect. \ref{sec:science}, the instrument concept, calibration, integration and simulated performance presented in Sect. \ref{sec:concept}-\ref{sec:performance} respectively. The summary and discussion are given in Sect. \ref{sec:discussion}.
	
	
\section{Science goals}\label{sec:science}
\subsection{Isotropic CXB measurement}\label{sec:sci_isocxb}
	
The CXB flux is roughly isotropic over the sky \citep{1987PhR...146..215B}. This isotropic flux could be averagely measured by an instrument able to collect X-ray photons from different fields (preferably blank sky) and filter out non-X-ray background \& known discrete sources. Previous measurements however remain affected by large uncertainties on the CXB spectral shape and normalization, which ultimately limit our knowledge of the accretion power and of the fraction of heavily obscured AGN in the Universe.
	
The CXB flux and spectrum have been measured by ASCA/SIS \citep{1995PASJ...47L...5G}, ROSAT \citep{1998A&A...334L..13M}, RXTE/PCA \citep{2003A&A...411..329R}, XMM-Newton \citep{2004A&A...419..837D}, Chandra \citep{2006ApJ...645...95H} and Swift/XRT \citep{2009A&A...493..501M} at soft X-rays, and by HEAO1 \citep{1980ApJ...235....4M, 1999ApJ...520..124G, 2005A&A...444..381R} and more recently by Beppo-SAX \citep{2007ApJ...666...86F, 1999A&A...349L..73V}, INTEGRAL \citep{2007A&A...467..529C, 2010A&A...512A..49T} and Swift/BAT \citep{2008ApJ...689..666A} at hard X-rays. The measurements are in agreement at a level of $10-15\%$ throughout the full energy range \citep{2005A&A...444..381R, 2007ApJ...666...86F, 2008ApJ...689..666A, 2009A&A...493..501M}.
	
The method to perform the CXB synthesis has been developed in the seminal works of \cite{1989A&A...224L..21S, 1995A&A...296....1C} and improved in the following works. To synthesize the CXB spectrum three main “ingredients” must be known: an accurate description of the broadband spectra of the various AGN classes (including reflection effects), the X-ray luminosity function (XLF) which gives the number density of AGN per comoving volume as a function of luminosity and redshift, and the distribution of AGN as a function of absorbing column density ($N_{\rm H}$), the so-called $N_{\rm H}$ distribution.
	
AGN spectra are provided as spectral templates for the various AGN classes: previous works set the parameters of their spectral templates to values representative of observations \citep[e.g.][]{2007A&A...463...79G} or of models \citep[e.g.][]{2014ApJ...786..104U}. \cite{2016A&A...590A..49E} used spectral templates derived by stacking Swift/BAT data for various types of AGN. The AGN X-ray luminosity function is derived from deep surveys \citep[e.g.][]{2003ApJ...598..886U, 2005ApJ...635..864L, 2005A&A...441..417H, 2010MNRAS.401.2531A, 2014ApJ...786..104U, 2015ApJ...804..104M, 2015ApJ...802...89B, 2015MNRAS.451.1892A, 2016A&A...590A..80R} and hence is available only in the soft X-ray range. The $N_{\rm H}$ distribution can be derived from data \citep[e.g.][]{2007A&A...463...79G}, but it is biased against the detection of highly absorbed sources) or from models \citep[e.g.][and references therein]{2009ApJ...696..110T}.
	
The XLF is directly proportional to the CXB normalization, while the spectral templates and the $N_{\rm H}$ distribution are mostly determined by the CXB spectral shape, in particular to the spectral slopes below and above \SI{30}{keV} and the break in between. An uncertainty of 10\%-15\% on the spectrum of the CXB at soft and hard X-rays corresponds to an uncertainty of up to 0.1 on the spectral slopes and of 0.2 on the spectral break. Considering that spectral templates derived from Swift-BAT have similar calibration as the BAT-derived CXB spectrum, the above uncertainties could be divided by a factor of two when considering such templates when comparing the CXB synthesis to the BAT observations.
	
The contribution of Compton thick (CTK) sources to the CXB flux at \SI{30}{keV} is estimated to be 4\%-6\% based on the CXB synthesis. As this is twice less than the 10\%-15\% uncertainty this contribution is highly dependent on CXB spectral uncertainties. A determination of the CXB spectral shape with a few percent levels of uncertainty is therefore required to better estimate the fraction of CTK AGN.
	
Our proposed detector attempts to measure the CXB spectrum in the 10-\SI{100}{keV} with 1\% precision, to significantly improve the study of the above topics. This instrument could also extend the CXB measurement up to \SI{1}{MeV} (although with less accuracy). Current data are scarce in this range and an improvement is of interest. The integrated Hawking radiation (HR) of primordial black holes (PBHs) with different mass scales could leave a signature in the isotropic CXB at energies above \SI{100}{keV}. \cite{2021PhRvD.103j3025I} and \cite{2021arXiv211215463C} derived upper limits to the PBH density using the diffuse X/$\gamma$ ray background. Extending the CXB spectral measurement to \SI{1}{MeV} together with a better estimate of the contribution of AGNs,  derived from lower energies, will improve significantly these limits \citep{2021PhRvD.103j3025I, 2021PhRvD.104b3516R}.
	
\subsection{Anisotropic CXB measurement}\label{sec:sci_anisocxb}
	
The CXB is characterized by small-scale fluctuation and large-scale anisotropy. The CXB is a superposition of numerous discrete sources including undetected/unresolved faint sources, while the numbers of them statistically fluctuate from field to field on small scale. This imprints fluctuations on the CXB fluxes with a scale of $\Omega_e^{-0.5} S_{\rm min}^{0.25}$ \citep{2012PASJ...64..113M}, where $\Omega_e$ is the field size for evaluation, and $S_{\rm min}$ is the detection threshold of the instrument. Detecting such fluctuations is out of the scope of our instrument, however, a dipolar anisotropy of the CXB is expected. First, our proper motion with respect to the distant Universe, where the bulk of the CXB is emitted, should result in a dipolar anisotropy with an amplitude of $0.42\%$ due to the Compton-Getting (CG) effect \citep{1935PhRv...47..817C}, in a direction matching that of the Cosmic Microwave Background (CMB) dipole \citep{2000ApJ...544...49S}. Then the distribution of AGNs in the local Universe could produce a small additional anisotropy, the amplitude of which has been estimated as $0.23-0.85\%$ by \cite{2000ApJ...544...49S}. HEAO-1 A2 \citep{1983IAUS..104..333S, 1987PhR...146..215B} and RXTE \citep{2008A&A...483..425R} have measured a dipole amplitude of $<$ 3\% and of $\sim$2\% respectively after subtraction of the CG effect. \cite{2008A&A...483..425R} further found that most of the observed CXB anisotropy (CG effect subtracted) can be attributed to low-luminosity AGNs. Better observational constraints are needed to improve these estimates. This requires however an accuracy of 0.1-0.5\% on the average CXB intensity.
	
\subsection{Secondary goal}\label{sec:sci_secondary}
	
Gamma-Ray Bursts (GRB) are among the most energetic explosions since the Big Bang, with an average rate of one event per day at cosmological distances \citep{2015JHEAp...7...73D}. Assuming an isotropic GRB occurrence,	we don't expect to see serendipitous GRB in the FoV of our instrument (only one in 5 years). However, the prompt emission of short Gamma-Ray Bursts (sGRBs) is generally hard with peak energies reaching $\sim$\SI{490}{keV} \citep{2011A&A...530A..21N}. Photons of such sGRBs could therefore penetrate the platform structure/instrument housing, and reach the detector. Missions like \textit{INTEGRAL} and \textit{Insight-HXMT} have successfully employed their inside anticoincidence detectors to monitor GRBs with nearly omnidirectional FoV \citep{2017ApJ...848L..15S, 2018SCPMA..61c1011L}. Based on the $\log N-\log P$ relationship (where P is the 50-\SI{300}{keV} peak flux) of the 4th BATSE GRB Catalog \citep{1999ApJS..122..465P} and the anticipated instrumental performance of our instrument (see Sect. \ref{sec:performance}, averagely $\SI{36}{cm^2}$ for off-axis effective area and background level of \SI{76.5}{cnts/s} in 100-\SI{300}{keV}), we expect roughly a detection rate of $\sim$4 GRBs (about 3 long and 1 short) per year with a detection significance $>5\sigma$. The brightest GRB(s) could be localized with an accuracy of a few degrees thanks to the directional dependent instrumental response \citep[see e.g.][]{2019ApJ...873...60B,2021NIMPA.98864866W}. 
	
Luminous gamma-ray pulsars like the Crab pulsar and PSR B1509-08 will be detected each time they pass in the FoV so they can be monitored over the duration of the mission. The timing resolution (at a level of $\mu s$) and the effective area will be good enough to detect their pulsations. Even when the Crab pulsar is out of the FoV, high-energy photons can penetrate the collimators and be detected using phase folding even in case of a very unfavorable signal over background level, as demonstrated with the POLAR detector \citep{2017ICRC...35..820L, 2019JHEAp..24...15L}. Detection of the pulsar(s) can be used to calibrate the absolute time stamping and to perform pulsar navigation \citep{2017ICRC...35..820L, 2017SSPMA..47i9505Z}. Dedicated simulations will be performed later to evaluate such secondary goals.
	

\section{Instrument Concept}\label{sec:concept}
	
\subsection{Overview}\label{sec:instrument_overview}
	
The diffuse nature of the CXB makes it difficult to be separated from the instrument background. Bright sources, additional sources of diffuse background and instrumental background need to be filtered out. An accurate method to distinguish these different components is crucial. 
	
Some experiments like ASCA/SIS \citep{1995PASJ...47L...5G},  Beppo-SAX \citep{2007ApJ...666...86F, 1999A&A...349L..73V} and RXTE/PCA \citep{2003A&A...411..329R} have deeply exposed some high galactic latitude blank sky regions to measure the CXB subtracting the same level of exposure obtained on the dark side of the Earth. INTEGRAL \citep{2007A&A...467..529C, 2010A&A...512A..49T} and Swift/BAT \citep{2008ApJ...689..666A} used Earth occultations, during which the Earth transits the field of view and modulates the CXB and other components. HEAO-1 \citep{1999ApJ...520..124G} used an onboard obturator to separate the CXB from other components. 
	
The detector proposed here utilizes passive collimators and onboard obturators to model the fluxes registered in the detector. Collimators (see Sec. \ref{sec:collimator} \& \ref{subsec:Lead_chamber}) are used to block surrounding emissions with energy-dependent transparency to reduce the contamination out of the FoV. Obturators (Sec. \ref{subsec:obturator_wheelsys}) will periodically shield the aperture of the collimator, and introduce a modulation of in-FoV components to separate them from the instrument background noise. To drive such obturators, a compact wheel system is developed (Sect. \ref{subsec:obturator_wheelsys}).
	
The science goals mentioned in Sect. \ref{sec:science} requires observing an energy range of 10-511 keV with a sensitive spectrometer (Sec. \ref{subsec:unit}). We propose to use a new generation of CeBr$_3$ scintillating crystals (Sect. \ref{subsec:crystal}) which have been studied to assess their suitability as spectrometer modules for space missions \citep{2004ISCRNS...7.4278, 2008ITNS...55.1391D, 2013NIMPA.729..596Q, 2016RScI...87h5112K, 2019JInst..14P9017D, 2021ExA....52....1M}. 
	
Overall, the detector consists of an array of collimated spectrometers with rotating obturators on top of the collimators. Hereafter, we present the concepts and mechanical designs of the aforementioned components of the detector.
	
\subsection{Collimator}\label{sec:collimator}
	
The collimator is a cylindrical tube made of four metal layers which are Aluminium (Al), Tin (Sn), Copper (Cu) and Al from outer to inner. These multi-layers of Al-Sn-Cu-Al shield off-axis X-ray photons. The innermost layer emit K-shell fluorescence ($<$ \SI{2}{keV}) below the energy threshold of the detector. With thicknesses of 1-1-1-\SI{2}{mm} for the Al-Sn-Cu-Al layers (the effective thickness is thicker by a projection factor of 1/$\sin \theta$, where $\theta$ is the incident angle), photons below $\sim$\SI{100}{keV} are expected to have a $<$0.1\% transparency through the collimator tube. The length and inner diameter of the tube is 250 mm and 25 mm respectively, which result in a FoV of around 26 square degrees (Full Width at Half Maximum, FWHM).
	
\subsection{Spectrometer}\label{subsec:unit}
	
\subsubsection{Crystal}\label{subsec:crystal}
	
The CeBr$_{3}$ crystal will be used as scintillation material of the spectrometer module, with a diameter and thickness of 25 mm and 20 mm respectively. The CeBr$_{3}$ provides improved detection performances with high light output (60 photons per keV), excellent energy resolution ($\sim 4$\% at \SI{662}{keV}, FWHM) and fast fluorescence decay time (17 ns) \citep{2013NIMPA.729..596Q}, which makes it suitable for the SiPM readout. With a thickness of \SI{20}{mm}, the CeBr$_{3}$ will keep $\sim$100\% detection efficiency up to \SI{200}{keV} and still reach a 60\% efficiency at \SI{511}{keV}, which is the emission line from a $\beta^+$ decay calibration source. \cite{2007NIMPA.572..785O} and \cite{2016RScI...87h5112K} have tested the CeBr$_{3}$ crystal with increasing proton fluences ranging from $10^9$ to $10^{12}$ $\SI{}{protons\ cm^{-2}}$, while the light yield was barely affected and the measured FWHM energy resolution at 662 keV was extended only by 0.1\%. This shows its high radiation hardness against proton-induced damage, which guarantees a stable performance without severe degradation in space. 
	
The CeBr$_{3}$ crystal will be sealed in a 0.1mm thick Aluminum frame with an internal optical reflector on the top and sides. The top will additionally be coated by a 0.1 mm thick beryllium window to stop low-energy charged particles. The resulting low energy threshold for X-ray photons will be $\sim\SI{10}{keV}$. At the bottom, an optical interface will connect the crystal to a Quartz window and to the Silicon photomultiplier.
	
\subsubsection{Silicon PhotoMultiplier}\label{subsec:SiPM}
	
Silicon PhotoMultiplier (SiPM) is increasingly used in space-borne detectors. It has large Photo Detection Efficiency (PDE), and its light yield lowers the low energy threshold. Other properties, such as good quantum efficiency, low bias voltage, compactness, robustness and insensitivity to magnetic fields, relax the detector design constraints. 
	
A disadvantage of SiPM is its high dark current, which is highly temperature-dependent and would increase as radiation dose accumulates in space. The temperature will have to be kept as low as possible, so that the temperature variation can be measured and calibrated. The radiation damage is more problematic and unavoidable, it will result in an increase in the threshold energy of the detector. The SiPM degradation was studied\footnote{\url{https://indico.cern.ch/event/1093102}} and constrained with an irradiation campaign \citep{proton..irradiation} showing that the CeBr$_3$ crystal is efficiently protecting the SiPM and that an order of magnitude of increase of the dark current can be expected after one year in space which would result in a threshold increase of a few keV per year at around $\SI{-20}{^{\circ}C}$ \citep{2021arXiv210900235Z, 2022arXiv220510506Z}.
	
\subsubsection{Electronics}\label{subsec:electronics}
	
A suitable electronic was already developed for the 64 channels SiPM array of POLAR-2 \citep{2022icrc.confE.580D}. A Front-End Electronics (FEE) board (CITIROC-1A ASICs) is equipped for each polarimeter module of POLAR to power (temperature regulated) and readout the SiPM channels, define the trigger logic, and communicate with a Back-End Electronics (BEE) board connected behind. The BEE board takes care of power supply, overall data acquisition, communication with the platform and mission control, etc.
	
\subsubsection{Bottom Lead chamber}\label{subsec:Lead_chamber}
	
A Lead (Pb) chamber will be additionally placed at the bottom to reduce the effective radiation doses received by the scintillators and SiPMs from the back side of the instrument. An 8-mm thick Pb is able to stop 100 keV photons at a level of more than 99\% percent.
	
\subsection{The obturator and wheel system}\label{subsec:obturator_wheelsys}
	
\subsubsection{Obturator}\label{subsec:obturator}
	
The obturator consists of the same sandwich layers as the collimator but twice thicker (\SI{10}{mm}, \SI{0.75}{kg}). There are two obturators configured as a propeller, which are counter-rotating to compensate the moment of inertia ($\SI{0.002}{kg\ m^2}$ for each). The rotation period of the obturator is defined as one rotation per minute (rpm), during which the space environmental background does not change significantly, thus the background levels registered in the tubes during the transit of the obturator are constant. An angular encoder is set in the wheel system to constantly record the rotation phase of the obturator.

\begin{figure}[!htbp]
		\centering
		\includegraphics[width=7cm,height=5.3cm]{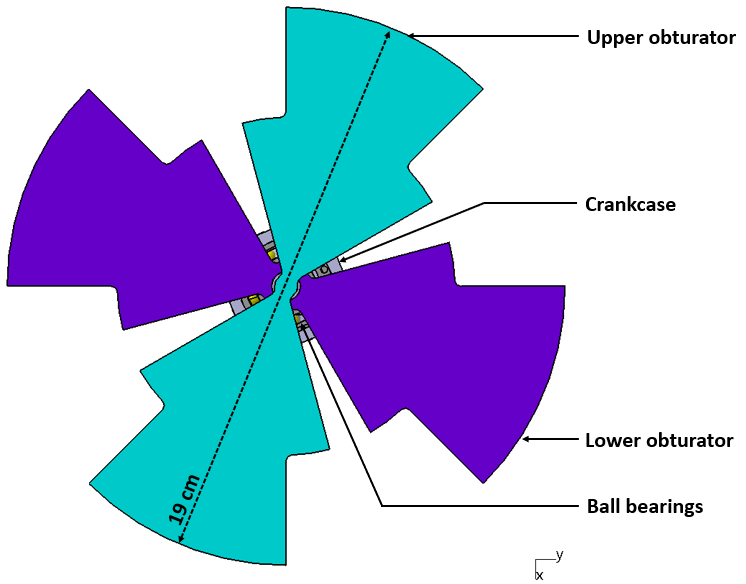}
		\caption{Schematic of the obturators (top view). Two symmetrical obturators are counter-rotating to each other. Opening angles at inner and outer sectors are adjusted to shield the collimator apertures at a different distance from the center.}
		\label{fig:propeller}
	\end{figure}

A schematic drawing of the obturator is shown in Fig. \ref{fig:propeller}. Each obturator is individually symmetrical to avoid shifting the center of mass. The opening angles of the inner and outer sectors are defined to be $\SI{75}{^{\circ}}$ and $\SI{45}{^{\circ}}$, such that there are always few tubes being closed by the obturators. Furthermore, the closures by one or two obturator(s) offer a chance to study the induced radioactivity of the obturator and collimator (similar layers) in space, thus helping to understand the internal background of the instrument.
	
\subsubsection{Wheel system}\label{subsec:wheelsys}
	
\begin{figure}[!htbp]
		\centering
		\includegraphics[width=9cm,height=4.5cm]{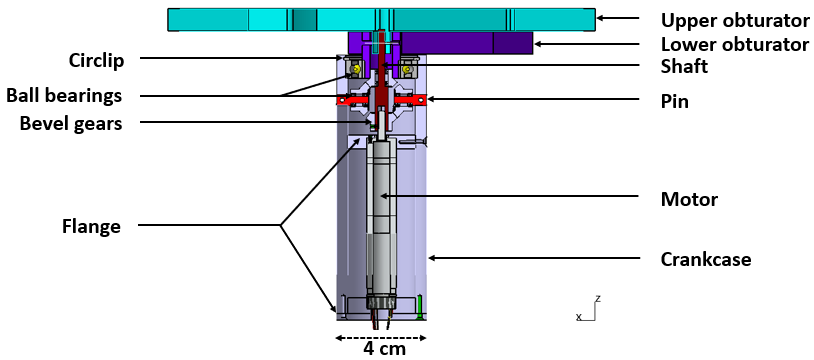}	
		\caption{Schematic of the wheel system (cutaway view). The obturators are driven by such a compact wheel system, which offers two counterrotating coaxial shafts from the unidirectional shaft driven by a motor.}
		\label{fig:wheelsys}
	\end{figure}

The obturators will be driven by a compact wheel system, which is adapted in a crankcase with a length of 150 mm and a diameter of 40 mm. The combination of the obturators and wheel system is mushroom-like, where the root will be inserted at the center of the collimator array (see Sect. \ref{subsec:cubesat}), such that the obturators could cover the aperture of the collimator tubes. Fig. \ref{fig:wheelsys} shows the cutaway view of the wheel system. The gearing provides two counterrotating coaxial shafts from the unidirectional shaft driven by the motor. A model of this complete system has been built using a Maxon motor\footnote{\url{https://www.maxongroup.net.au/maxon/view/content/overview-BL-DC-Motoren}} and underwent some preliminary space qualification. Tagged radioactive sources will be attached beneath the lower obturator (see Sect. \ref{subsec:orbit}). The wiring of the tagged source requires standard slip rings in the wheel system for signal connection.
	
Moving mechanisms working in space are difficult. First, the gears should be robust and especially the ball bearings have to overcome the grouping, vibration and frictional torque problems as shown in \cite{2001ESASP.480..289O, 2013ESASP.718E...8V}. Then the motor is required to work in a vacuum and in a large temperature regime \citep[see e.g.][]{Phillips2012DevelopmentOB, maxonnews}. Thermal-vacuum tests need to be carried out for space qualification to reduce risks of failure. In the worst-case scenario if the propeller is stopped, the detector loose flux modulation, but the partially or entirely closed and open tubes still enable background modeling.

\subsection{Thermal constraints}\label{subsec:thermal}
	
In order to reduce the impact of dark noise of the SiPM and to maintain a relatively low energy threshold, a cooling system is needed to reach ideally below $\SI{-20}{^{\circ}C}$. The solar radiation power to the instrument is orbit dependent as the Sun incident angle varies. The averaged power over one year can be estimated at $\SI{1361}{W m^{-2}}$ beyond the Earth's atmosphere \citep{2016BAMS...97.1265C}, and the planetary albedo at about 35\% \citep{2015RvGeo..53..141S}. The whole instrument with dimension given in Sect. \ref{subsec:cubesat} will heat by an average value of \SI{13}{W}. Taking into account $\sim$\SI{30}{W} of additional heating from the electronics, the total cooling power needed is $<$\SI{43}{W}. The tubes of the instrument (with a total area of $\SI{0.21}{m^2}$ for 18 tubes) could serve as passive radiators for dissipation. The calculated thermal equilibrium of the instrument is expected to be \SI{252}{K} ($\SI{-21}{^{\circ}C}$) with several degrees of variation because of the Sun power modulation along a 90-minute orbit.

\section{Calibration}\label{sec:calibration}

Accurate knowledge of the detector's spectral properties, including the Energy-Channel (E-C) relationship (linearity), energy resolution, detection efficiency, etc., is crucial to measure the CXB normalization. Some of the calibrations will be carried out on ground and some need to be performed during operations.
	
The absolute detection efficiencies (versus energy and time) depend on geometry (FoV), photopeak versus Compton ratio, light yield of CeBr$_3$, scintillation light efficiency of transmission and collection, the quantum efficiency of the SiPM, etc. The geometry needs to be characterized very precisely on the flight model before launch and is not expected to change afterwards. During the mission, gradual changes of some parameters are expected: i) the SiPM performance varies with temperature but this can be adjusted (biasing voltage on SiPM is automatically adjusted at all times by the electronic); ii) the SiPM performance can change due to radiation damage; iii) all the CeBr$_3$ characteristics can change due to aging and radiation damage. The latter two will have an impact on the E-C relationship, energy resolution and detection efficiency, which need to be monitored in orbit.
	
\subsection{On ground}\label{subsec:ground}
	
Before launch, the detector will be fully assessed and its spectral responses measured using standard radioactive sources, such as $^{241}\rm Am$ (half-life 432.2 years, specific activity 126.8 GBq/g) and $^{22}\rm Na$ (2.6 years, 1580 TBq/g). The $\alpha$ decay of $^{241}\rm Am$ generates gamma rays of 13.9, 17.8, 26.4 and \SI{59.6}{keV}. $^{22}\rm Na$ has $\beta^+$ decay emitting a positron immediately annihilating and releasing two gamma-ray photons at \SI{511}{keV} in coincidence with a 1.2 MeV photon. The energy-channel relationship and energy resolution at different energy peaks will be obtained by measuring the channel spectrum of these sources with the detector.
	
The absolute detection efficiencies versus energy can be measured by recording the single and coincident photon counts of tagged radioactive sources peaking at different energies. For example, the decay of a $^{241}\rm Am$ source, continuously monitored by a tagged scintillator, can be marked by the triggering of the scintillator above a relatively high threshold ($\alpha$ decay at MeV). Then a detector module can be placed near the source to collect photons. The expected number of photons that should arrive at the detector is in direct proportion to the counts of the scintillator by a ratio of collecting solid angle over the allowed emission angle ($4 \pi$ assuming an isotropic decay). Eventually, the division of the coincident photon counts to the expected number gives efficiency.  The accuracy of the efficiency calibration is purely limited by statistics, which is approximately the square root of one over the coincident counts. At \SI{511}{keV}, an alternative calibration approach is to use the $^{22}\rm Na$ source, which can be placed between two detector modules, then the single and coincident events offer an equation set to resolve the unknown efficiencies of both detector modules.
	
Finally, a synchrotron radiation facility could generate a collimated broadband X-ray beam with precisely known spectra. An X-ray continuum (covering 10-\SI{100}{keV}) irradiating the instrument, allows to calibrate the spectral response as a function of energy, cross-checking and filling the gaps continuously between the discrete energies provided by the radioactive sources. 
	
	

	%
	
	

A Monte-Carlo simulator of the instrument based on Geant4 \citep{2003NIMPA.506..250A} will be developed to verify all the aforementioned calibration procedures. Additionally, as the thermal conditions will change in orbit, we will evaluate all the temperature-regulated parameters, in particular these of the SiPM and electronics. The degradation of the SiPM and of the crystal to radiation dose will be evaluated through irradiation with protons (the major background in space, see \ref{subsec:background} for details).
	
\subsection{In orbit}\label{subsec:orbit}
	
There are two specific challenges in orbit. The first is to cross-check all the parameters calibrated on-ground. Secondly, the detectors must be monitored for aging and degradation due to radiation. To cope with them, there will be two tagged $^{241}\rm Am$ sources attached beneath the lower obturator (as shown in Fig \ref{fig:CubeSat}, both the inner and outer sectors of the lower obturator will be attached with one source, which will pass by the center of the inner and outer tubes). As the calibration sources are periodically transiting in the FoV, the calibrations can be done independently for every detector tube (they should be uniform at the beginning). The first in-orbit calibration will validate the spectral response characterized on ground. 
	
The E-C relationship and energy resolution can be calibrated in the same way in orbit as on ground (one energy point is sufficient). This will be done constantly to monitor the aging and degradation of the detector. The E-C relationship is closely linked to the performances of SiPMs, which are sensitive to radiation dose and temperature. Their influence will be periodically checked. The energy resolution is anticipated to not be dramatically changed as the CeBr$_3$ has high radiation hardness (Sect. \ref{subsec:crystal}). Even a slight change will be monitored as time-dependent responses for scientific analysis.
	
The detection efficiency calibration in orbit is slightly different than on ground, since the latter is static whereas it is evolving in orbit. The rotation phase of the obturator allows bin the exposure of the calibration sources to every tube. The detection efficiency for every tube can then be calibrated as it was done on ground (\ref{subsec:ground}). 
	
A \SI{200}{Bq} $^{241}\rm Am$ source will create about $10^4$ events per day in the photopeak region of the \SI{59.6}{keV} photon of the tagged spectrum of each tube (0.99 tagging efficiency, 0.36 branching ratio of the decay at \SI{59.6}{keV}, 0.031 solid angle ratio, 0.13 effective exposure time, 0.6 photopeak efficiency, 0.8-1 deadtime ratio). So statistically the efficiency at \SI{59.6}{keV} of each tube can be measured at 1\% level each day. Some other lines of $^{241}\rm Am$ can be used to calibrate the efficiency, energy-channel relationship and energy resolution at lower energy but with a lower accuracy limited by statistics. Besides, an induced instrumental background will include a \SI{511}{keV} feature in the accumulated spectra. This line can also be used for calibrating the E-C relationship and energy resolution. Furthermore, since inner and outer detector tubes have different radiation acceptance, the evolution of their correlated spectral properties will allow to further characterize their degradation.
	
The selection of coincident events can be done offline or in real-time with a coincidence time resolution of 100 ns, to reduce telemetry. We expect 3 random coincident events per day and detector, which is completely negligible compared to the $10^4$ real coincidences. 
	
These calibrations will be performed every day to monitor aging and radiation dose. Additionally, the Crab broadband spectrum will allow to cross-check and monitor the shape of the energy responses of the instrument. Geant4 simulator will model the Crab observations with respect to the on ground calibrations and allow to model the gradual change of the spectral response.
	
\section{Detector integration and platform requirements}\label{sec:integration}
	
As the space environmental background is highly orbit-dependent (Sect. \ref{subsec:background}), a suitable orbit is important to achieve the science goals. The best orbit is an equatorial orbit that never enters the South Atlantic Anomaly (SAA), but flight opportunities are not frequent. The very popular Sun-synchronous orbit for CubeSat is not well suited for our detector as it passes frequently in the SAA and suffers from continuous Solar radiation. 
	
A typical low Earth orbit (LEO) has an altitude of 300-\SI{500}{km} and an inclination of a few tens degrees. Lower orbits ($<$\SI{400}{km}) are preferable but the payload would deorbit rapidly. \SI{500}{km} is the minimum altitude for a free flier without propulsion. We will assume in the following orbital altitude of \SI{500}{km} and inclination of $\SI{42}{^{\circ}}$. Since our instrument need to continuously scan outer space, pointing to the zenith is required. 
	
As the mission platform will constrain the available resources (size, mass and power consumption) and determine the performance of the detector, we considered a 12U CubeSat implementation and a station-based modular design.
	
\subsection{CubeSat version}\label{subsec:cubesat}
	
\begin{figure}[!htbp]
		\centering
		\includegraphics[width=8cm,height=7cm]{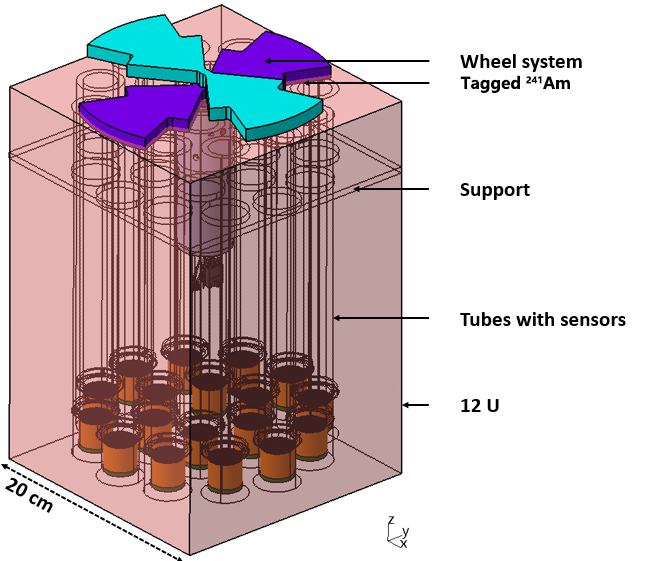}	
		\caption{Schematic of the CubeSat integration. The pink box symbolizes 12 U as a reference. the wheel system together with the obturators is at the top. 18 detector tubes are placed (6 inner and 12 outer) in the box. Two tagged $^{241}$Am sources will be attached beneath the bottom obturator for calibrations (both the inner and outer sectors of the lower obturator will be attached with one source, which will pass by the center of the inner and outer tubes).}
		\label{fig:CubeSat}
	\end{figure}

We have integrated our detector into a 12-Unit (U) CubeSat payload, translating to 2*2*3 U (one U corresponds to 10*10*10 $cm^3$). Such a configuration is shown in Fig. \ref{fig:CubeSat}, where the transparent pink box symbolizes 12 U as a reference. It allows the placement of 18 tubes (each includes a collimator and a spectrometer), all of which have the same dimensions: \SI{28}{cm} in height and \SI{35}{mm} in external diameter. 
	
The tubes are placed along a dual-ring structure. The 6 inner tubes will be shaded by the 12 outer ones and get less background from the sides, allowing to study the systematics of the background. A compact electronics (Sect. \ref{subsec:electronics}) will be placed beneath the wheel system in the center to power and communicate with the motor and spectrometers, the total power consumption is $\sim$\SI{30}{W}. 
	
The necessary platform modules (e.g., power, communication and orbit control)  could occupy the corners and sides or another 4-U. This very conservative configuration with 18 tubes is chosen to bring a lot of redundancy to fight systematic effects. A smaller number of tubes (as low as 4 tubes as seen in Fig. \ref{fig:measure_precision}) is possible if considering only statistical effects. Smaller CubeSat (4U, 8U) could probably reach the scientific goals if sacrificing redundancy but could require longer exposure to understand systematical effects due to the space environment. 
	
\subsection{Station-based version}\label{subsec:station}
	
\begin{figure}[!htbp]
		\centering
		\includegraphics[width=8cm,height=7cm]{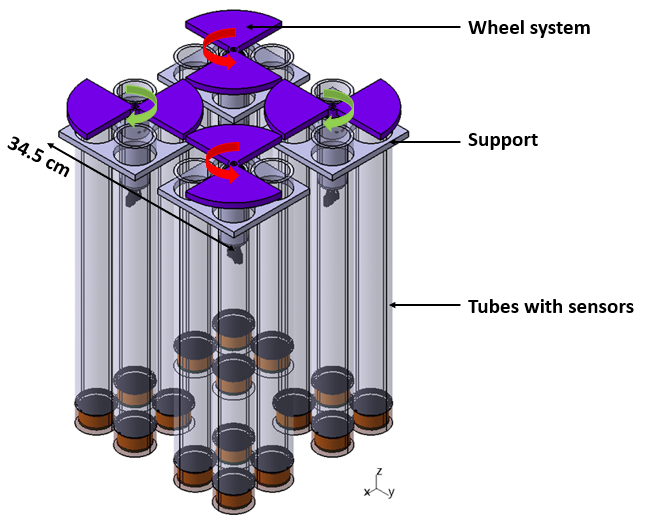}	
		\caption{Schematic of the Station-based integration. There are four groups, each containing four detector tubes which are twice the size of that in the CubeSat version. Each obturator is one symmetrical sector and is driven inversely by a unidirectional wheel system.}
		\label{fig:Station}
	\end{figure}

Another configuration for a Station-based platform is shown in Fig. \ref{fig:Station}. It is based on four groups, each containing four tubes that are twice the size of the CubeSat version, with a height of \SI{500}{mm} and an inner diameter of \SI{50}{mm}. This provides a collecting area four times bigger than of the CubeSat version. Four groups of obturators are placed on the top, each including a symmetrical sector (opening angle is $\SI{90}{^{\circ}}$). Neighboring obturators are counter-rotating to compensate for angular momentum. The wheel system is simplified as it needs unidirectional rotation.
	
\subsection{Technical Readiness}\label{sec:readiness}
	
The readout electronics are expected to be ready in early 2023 by adapting that of POLAR-2. A prototype of the detector unit will be built right after, to characterize its spectral response and detection efficiency. The overall design will then be finalized, integrating a test model of the detector for comprehensive qualification tests including motor, thermal, vibration, radiation and geometry, lasting until the end of 2023. As soon as a launch opportunity is determined, a flight model will be built by integrating with the platform interfaces and tested. Overall, a flight model can be expected to be delivered in 24-36 months.
	
\section{Simulated performance}\label{sec:performance}
	
\subsection{Detector spectral responses}\label{subsec:responses}
	
In this section, we consider the CubeSat configuration (Sec. \ref{subsec:cubesat}) to evaluate the performance (the station-based version (Sec. \ref{subsec:station}) has better statistics thanks to a bigger photon collecting area). The spectral responses and the background are generated using the Geant4 simulation package (\textbf{version 10.6.2}, \cite{2003NIMPA.506..250A}). Geant4 integrates comprehensively the relevant physics processes.
	
\begin{figure*}[t]
		\centering
		\includegraphics[width=13cm,height=5cm]{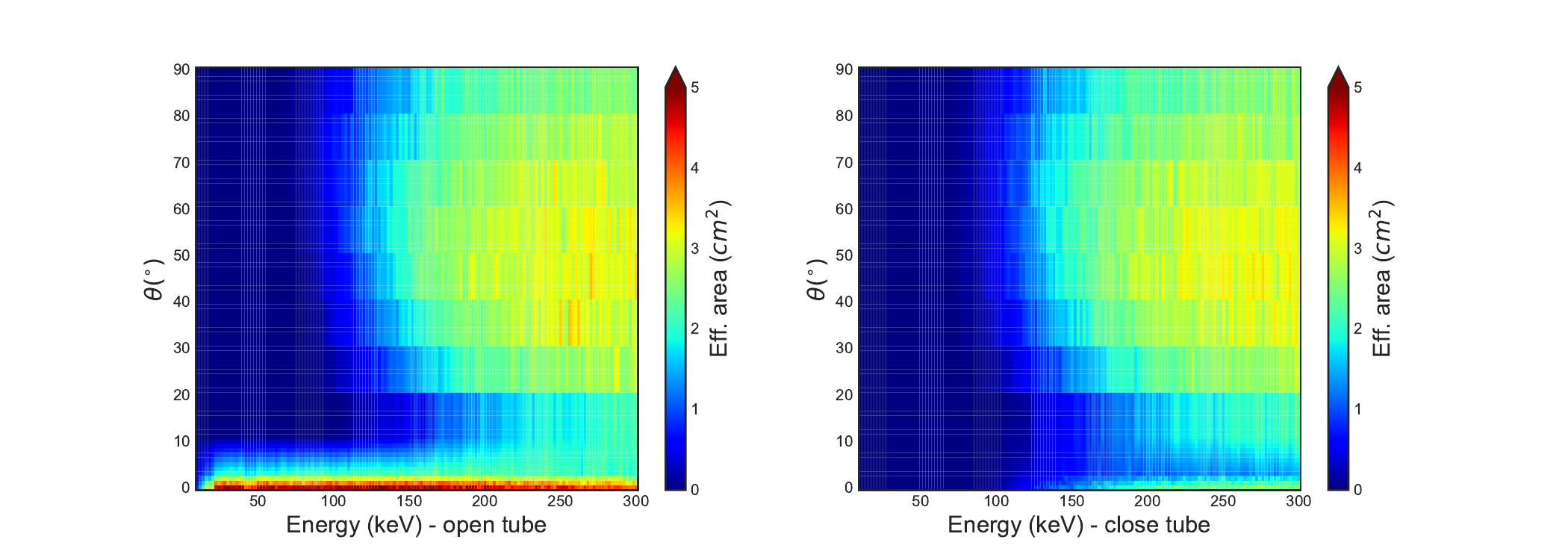}	
		\caption{Effective area for open/close tubes: the X-axis is the photon energy and the Y-axis is the incident theta angle. The left and right panels are for an open and closed tube respectively. One can see the opening and closing by the obturator and collimator work well for photons ranging from 10-100 keV, which offer a way to discriminate photons coming out-FoV (incident zenith angle bigger than $\sim\SI{6}{^{\circ}}$).}
		\label{fig:arf}
	\end{figure*}

The mass model of a single tube unit (obturator, collimator, and spectrometer) and the related physical processes have been implemented in Geant4. Monoenergetic photons (covering 10-\SI{1000}{keV} with reasonable steps) have been injected with different incident angles and shield coverages of the obturators. Fig. \ref{fig:arf} shows the effective areas for open and closed tubes as a function of incident angle $\theta$ and energy. The right panel indicates that the low energy threshold of the close tubes on-axis is $\sim$\SI{100}{keV}. The left plot indicates that the opening angle of the open tubes below \SI{100}{keV} is $\sim\SI{6}{^{\circ}}$. The energy response matrices are also obtained as a function of incident angle and energy.
	
Spectral responses for the full detector (made of 18 tubes) were also calculated. Tubes of the inner ring are shielded by those of the outer ring, resulting in a non-uniform (or more directional dependent) response. The instrument has more triggers on the external sides of the outer rings tubes, anticipating a localization ability to luminous transient sources like GRBs. Such an idea has been widely applied by multiple instruments (e.g., we have done that for POLAR \citep{2021NIMPA.98864866W}), and will be developed for the instrument presented here in the future. A conservative estimation of the localization accuracy will be a few degrees. While for the CXB, the instrument response remains symmetrical along the zenith. 

\subsection{Expected count rate}\label{subsec:countrate}
	
The Burst Alert Telescope (BAT) onboard the Swift observatory has successfully carried on an all-sky hard X-ray survey at 14-\SI{195}{keV} and detected 1632 sources in 105 months \citep{2018ApJS..235....4O}. The source catalog is available on-line\footnote{\url{https://heasarc.gsfc.nasa.gov/W3Browse/all/xray.html}}. We used that catalog to predict the expected count rates from the sources. The CXB count rate was calculated by convolving spectral templates \citep{1999ApJ...520..124G} with the simulated detector responses. 
	
\begin{figure}[!htbp]
		\centering
		\includegraphics[width=8cm,height=6cm]{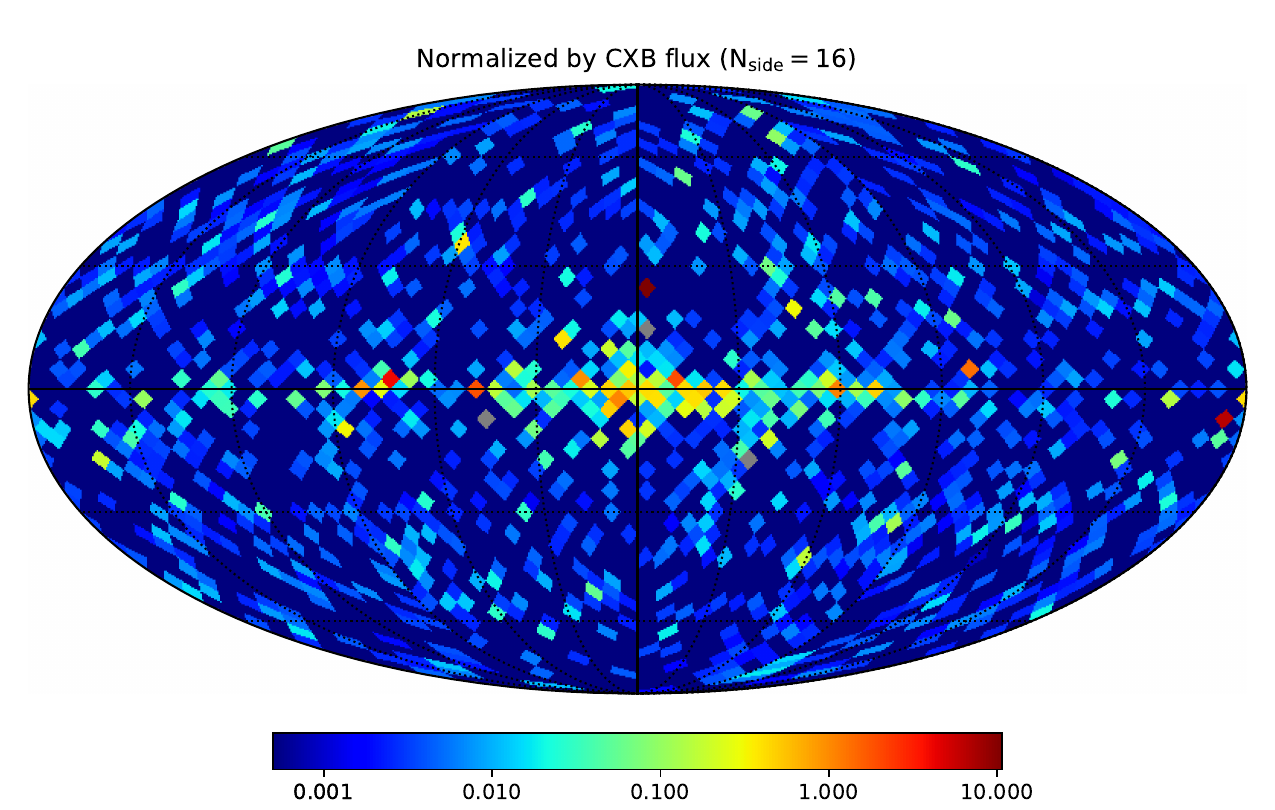}
		\caption{Distribution of the relative count rate of the Swift-BAT source catalog \citep{2018ApJS..235....4O} in our instrument (normalized by the CXB rate, i.e., 0.129 counts/s/tube). A cut in galactic latitude could reduce the majority of the luminous galactic sources, and thus improve the signal-to-noise ratio to the CXB measurement.}
		\label{fig:src2cxb_rate}
	\end{figure}

The derived count rate (10-\SI{100}{keV}) distribution over the sky is shown in Fig. \ref{fig:src2cxb_rate} in the unit of the CXB rate (0.129 counts/s/tube) with a bin size corresponding to the FoV of a tube. On average one source is contributing per sky bin. Many bright point sources on the galactic plane, could easily be filtered out. 
	
\begin{figure}[t]
		\centering
		\includegraphics[width=8cm, height=6cm]{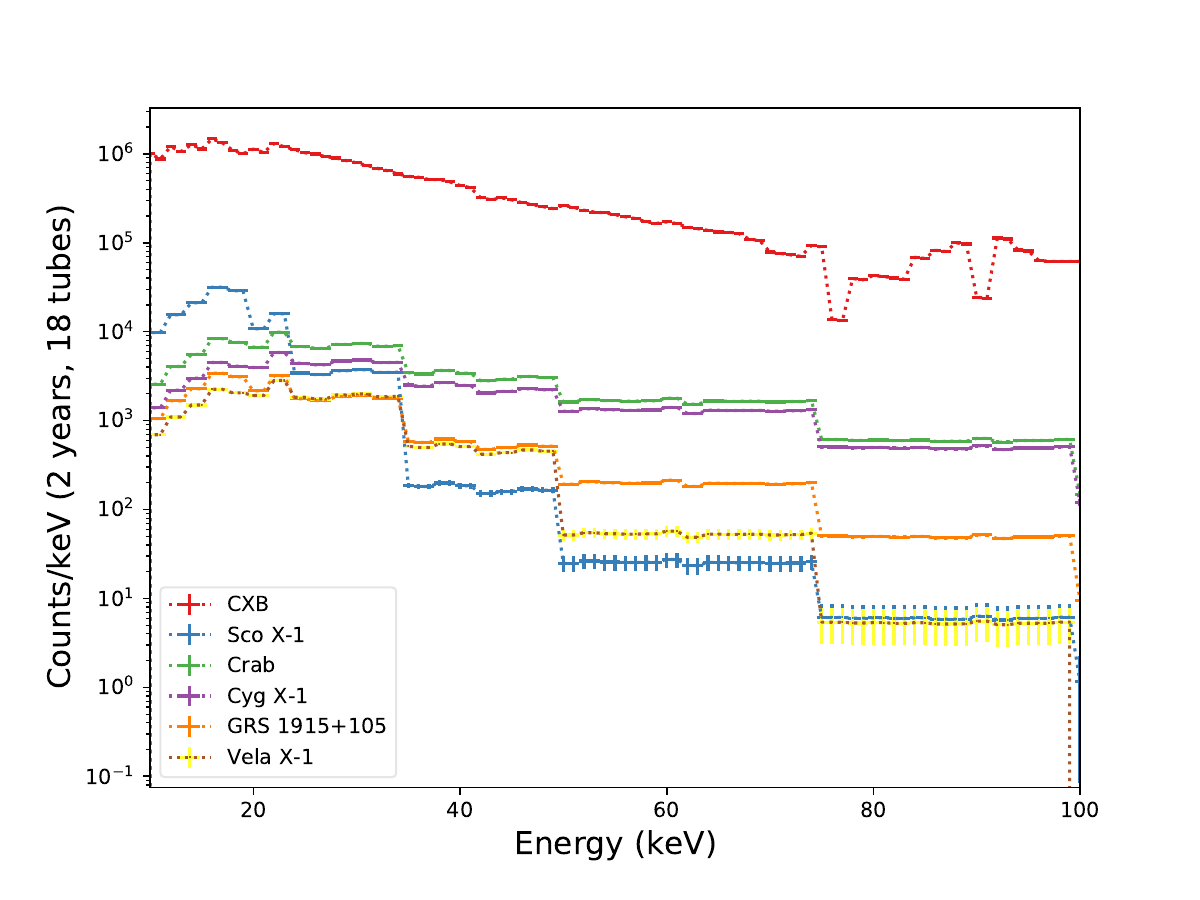}
		\caption{Counts spectra of the CXB and 5 luminous sources for two years by the CubeSat mission (18 tubes). The red curve represents the counts of the CXB. The others from top to bottom represent the counts of Sco X-1, Crab, Cyg X-1, GRS 1915+105 and Vela X-1 respectively.}
		\label{fig:partsources_flux}
	\end{figure}

We also plot the count spectra of the CXB and of five luminous sources in Fig.\ref{fig:partsources_flux}, for a 2-year exposure time and a CubeSat mission with 18 tubes (note that only the open tubes, i.e. half of them, contribute to these counts). The CXB will be detected with about 100 times more counts than the brightest sources (which could easily be excluded).
	
\subsection{Background estimation}\label{subsec:background}
	
We assume that the instrument will fly in a Low-Earth Orbit (LEO), where the space environmental background is the main concern. In our background studies, we have included the standard Shielding Physics List in Geant4, including electromagnetic physics, hadronic physics and radioactive decay physics (delayed background).

\subsubsection{Cosmic rays}\label{subsec:cr}
	
When approaching the Earth, low-energy cosmic rays (CRs) are deflected by the geomagnetic field. Higher-energy ones interact with the atmosphere creating  secondary particles creating noise in the detector. The South Atlantic Anomaly (SAA), caused by the offset of the Earth’s magnetic center from the geographic center, features a weaker geomagnetic field and an increased particle (mainly protons and electrons) environment \citep{zombeck2006handbook}.
	
Even if the instrument will be switched off in the SAA to protect the detector, the delayed background originating from the radioactivity induced by the SAA particles must be taken into account. A light flying unit (single CubeSat) will develop less induced background than that e.g. on a space station. 
	
We estimated the primary particle background spectra, based on the works by \cite{2018Galax...6...50X} and \cite{2022APh...13702668Z}, for an orbit at a typical altitude of \SI{500}{km} and inclination of $\SI{42}{^{\circ}}$. The majority of the primary CRs are protons with a spectrum that can be approximated by a power-law \citep{2000PhLB..490...27A,2015PhRvL.114q1103A}. We used an orbital averaged spectrum extracted from ESA's SPace ENVironment Information System \footnote{\url{https://www.spenvis.oma.be}} (SPENVIS). Other significant particles are electrons and positrons and we used a spectral model developed by \cite{2004ApJ...614.1113M}, often used for background simulation of space instruments. 
	
The spectra of secondary particles depend on the geomagnetic latitude \citep{2004ApJ...614.1113M}. By averaging results obtained by AMS-01 \citep{2000PhLB..472..215A}, an averaged spectra of secondary protons could be extracted.	The spectra of secondary electrons and positrons are provided by \cite{2008AdSpR..42.1523G}. Only a very small portion of them could be mistaken as photons as the low-energy ones are stopped by the beryllium window and cannot reach the detector. 

\subsubsection{Albedo gamma rays}\label{subsec:albedo_gamma}
	
The Earth's atmosphere features outwards radiation of albedo gamma rays, produced either by the reflection of the CXB (minor) or by the interaction between CRs and the atmosphere. Their spectra are provided by \cite{2013ExA....36..451C}. Since they are coming from below the instrument, their influences are highly correlated to the angle between the Earth and the FoV and this can be used to filter them out. Those photons are rather soft and only the high-energy ones will be able to go through the tube shield. 
	
\subsubsection{Delayed background}\label{subsec:delayed_bkg}
	
The delayed background originates from the radioactivity mainly induced by the trapped particles in the SAA. Geant4 is able to simulate the amount of induced radiation hitting the instrument from the fraction of the time spent in the SAA. The radioactive isotope production and decay properties can be characterized. The level of delayed background increases with time and gradually saturates. This background will develop characteristic lines (mainly 511 keV). The material of the detector and of the platform has to be chosen to minimize radiogenic materials.
	
\subsubsection{Background rates}\label{subsec:bkg_rates}
	
\begin{figure}[h]
		\centering
		\includegraphics[width=8cm, height=6cm]{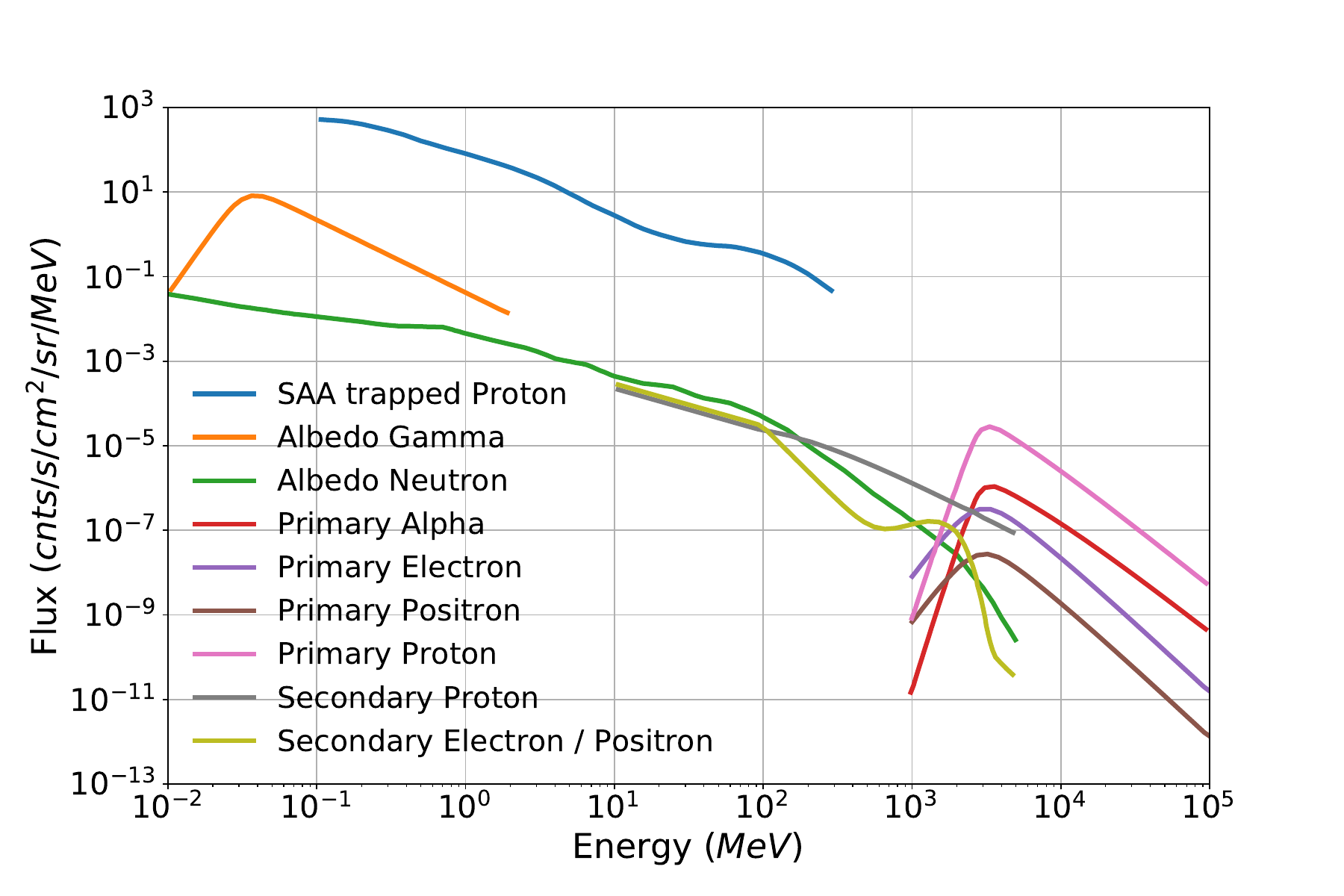}
		\caption{Input spectra of different background components for the Geant4 simulation, ranging from $10^{-2}$ to $10^{5}$ MeV. The data points are obtained from \cite{2018Galax...6...50X}, \cite{2022APh...13702668Z} and references therein.}
		\label{fig:bkg_simulation_input}
	\end{figure}

\begin{figure}[h]
		\centering
		\includegraphics[width=10cm, height=6cm]{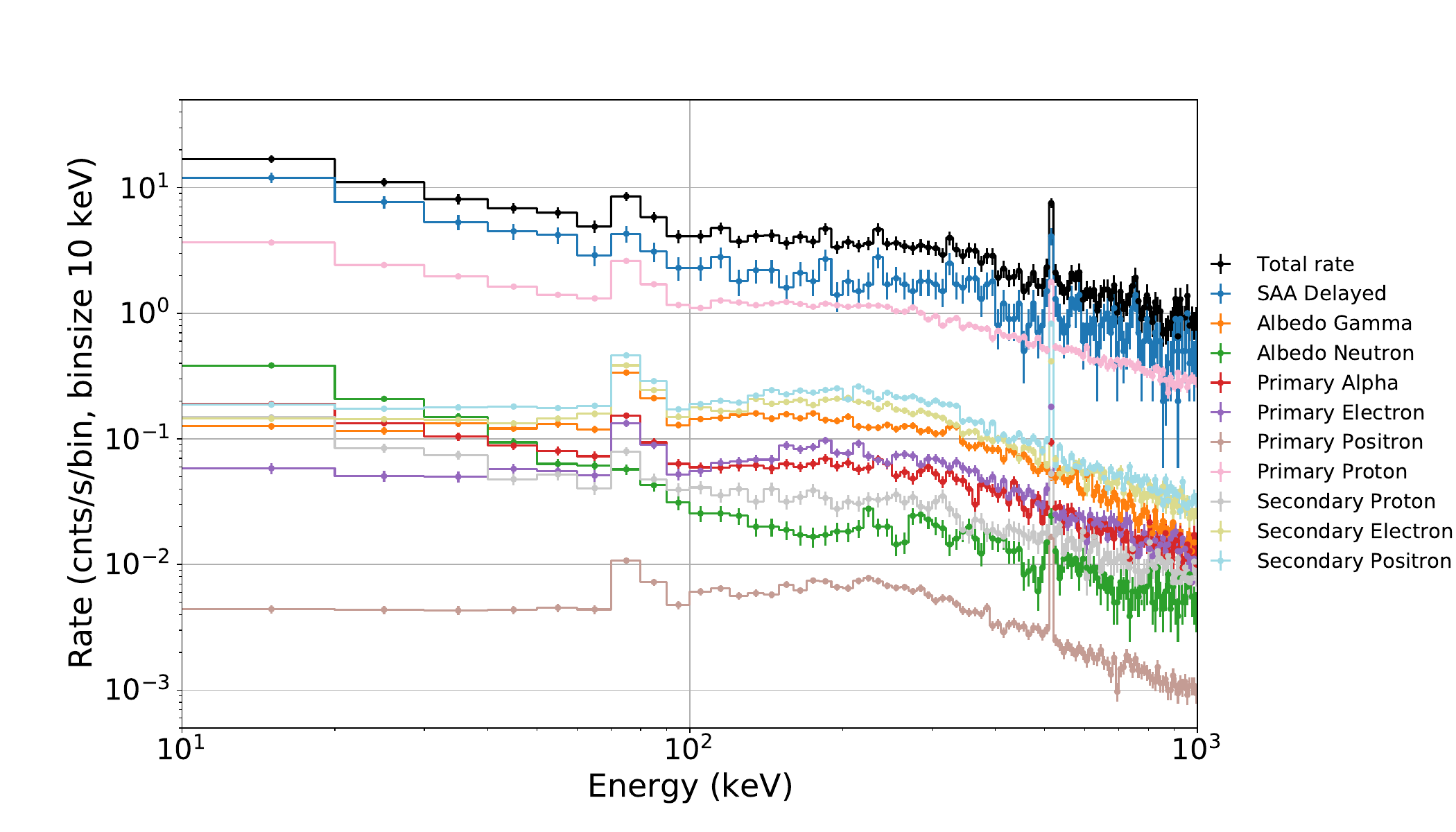}
		\caption{Simulated spectra of different background components. The mass model of the Geant4 is constructed by taking into account the CubeSat geometry defined in Sect. \ref{subsec:cubesat}. The total rate of 72.56 cnts/s in 10-100 keV is used in this paper for the estimation of measurement accuracy of the CXB.}
	\label{fig:bkg_simulation_output}
	\end{figure}

The spectra of the different components aforementioned, either adapted from the AMS measurements or from SPENVIS, are shown in Fig. \ref{fig:bkg_simulation_input}. They are all considered to be isotropic inputs to the instrument simulator, the mass model of which is constructed by taking into account the geometry defined in Sect. \ref{subsec:cubesat}. The anticipated background rates are shown in Fig. \ref{fig:bkg_simulation_output}. Table \ref{tab:bkg_rates} lists the rate of different background components in two energy ranges.
	
	\begin{table}[h!]
	\caption{Count rates (in the unit of cnts/s) of different background components in two energy bands for an 18-tube CubeSat configuration (half open half close). Primary cosmic rays in this table include primary alpha, electron, positron and proton; secondary cosmic rays include secondary electron, positron and proton; SAA (delayed) is the delayed radioactivity induced by SAA protons.}
	\label{tab:bkg_rates}
	\centering
	\begin{tabular}{ l c c  }
			\hline\hline
			& 10-100 keV (cnts/s) & 100-1000 keV (cnts/s)\\
			\hline

			Primary CRs & 19.49 & 64.05 \\
			Secondary CRs & 4.26 & 20.09 \\
			Albedo Neutron & 1.09 & 1.07 \\
			Albedo Gamma & 1.42 & 5.95 \\
			SAA (delayed) & 46.30 & 105.50 \\
			\hline
			Total background& 72.56 & 196.66 \\
			\hline
			CXB & 1.16 & 0.14 \\
			\hline
		\end{tabular}
	\end{table}

\subsection{CXB measurement}\label{subsec:CXB}

We are evaluating here the accuracy of the CXB normalization determination in the energy band 10-100 keV for an increasing number of tubes and exposure time. 
	
\begin{figure}[!htbp]
        \centering
        \includegraphics[width=7cm, height=6cm]{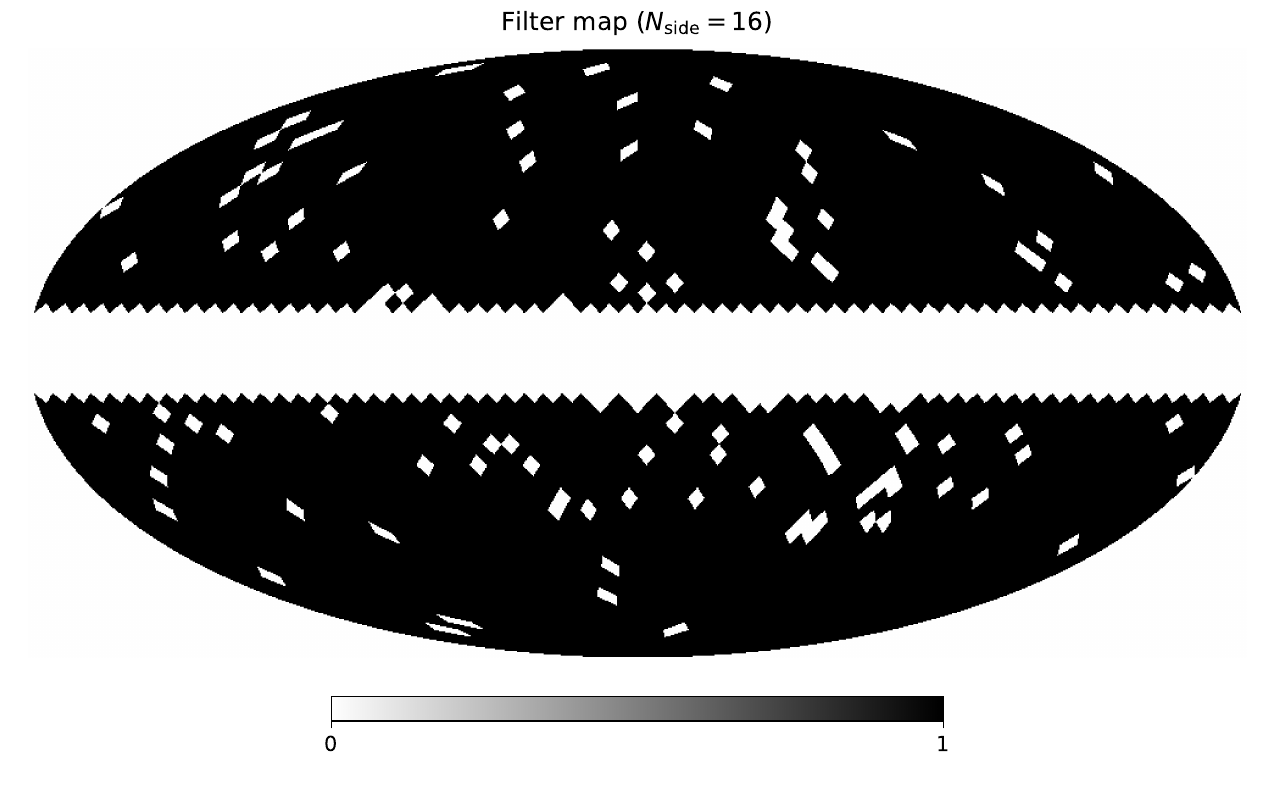}
        \caption{Filter map in galactic coordinates. The white pixels will be disregarded in the analysis (globally about 23\% of the whole sky). The pixels close to the galactic poles are contaminated by galaxy clusters and unidentified hard X-ray sources (likely AGN).}
        \label{fig:pixel_selection}
\end{figure}

For data selection purposes the sky is divided into 3072 pixels (healpix binning with $N_{\rm side} = 16$). Sky pixels with $|b|<10^{\circ}$ or covering the Magellanic clouds or including (non-AGN) sources in the Swift-BAT catalog \citep{2018ApJS..235....4O} are excluded. Most remaining sources at high galactic latitude are nearby CVs, stars and unidentified sources, which are the bright components of the GRXE population. Pixels with a count rate significantly above their neighbor will also be excluded. As shown in Fig. \ref{fig:pixel_selection}, globally about 23\% of the sky will be disregarded. 
	
Photons from unresolved galactic X-ray sources and from the instrumental background need to be subtracted from the observations. The statistical accuracy of the CXB measurement can be estimated as $P_{\rm sta} = (\sqrt{C+B}+U)/C$, where $C$, $B$ and $U$ represent the total number of counts detected from the CXB, the instrumental background, known galactic X-ray sources and of the unresolved Galactic Ridge X-ray Emission (GRXE) in all open tubes from all the considered sky pixels. 
	
The GRXE is made of numerous unresolved faint sources like coronally active binaries, cataclysmic variables, etc., and extends up to a galactic latitude of $|b|=\SI{20}{^{\circ}}$. It has a constant spectral shape, softer than the CXB, and an intensity (measured by INTEGRAL) scaling with the stellar mass \citep{2007A&A...463..957K} and in particular with the COBE/DIRBE Zodi-Subtracted Mission Average Maps at $\SI{4.9}{\mu m}$ \footnote{\url{https://lambda.gsfc.nasa.gov/product/cobe/dirbe_zsma_data_get.html}}. Using this map for the non-discarded pixels, the GRXE spectral template and convolving with the response of the detector indicates $U\sim 0.291$ counts/s/tube in total for $10^{\circ} < |b| < 20 ^{\circ}$, which correspond averagely to 0.47\% of the CXB rate. The GRXE count rate is very low ($<$ 0.067 count/s/tube) for $|b|>\SI{20}{^{\circ}}$ allowing a very good separation of the CXB and of the GRXE fitting the full sky.

As the detector tubes are switching from open and close state thanks to the obturators, the instrument background (about $B\sim 4.0$ counts/s/tube in the range of 10-100 keV, see Sec. \ref{subsec:bkg_rates}) can be determined and its evolution precisely modeled (at least within 0.1\% in the band 10-\SI{100}{keV}). 

\begin{figure}[t]
		\centering
		\includegraphics[width=10cm,height=7cm]{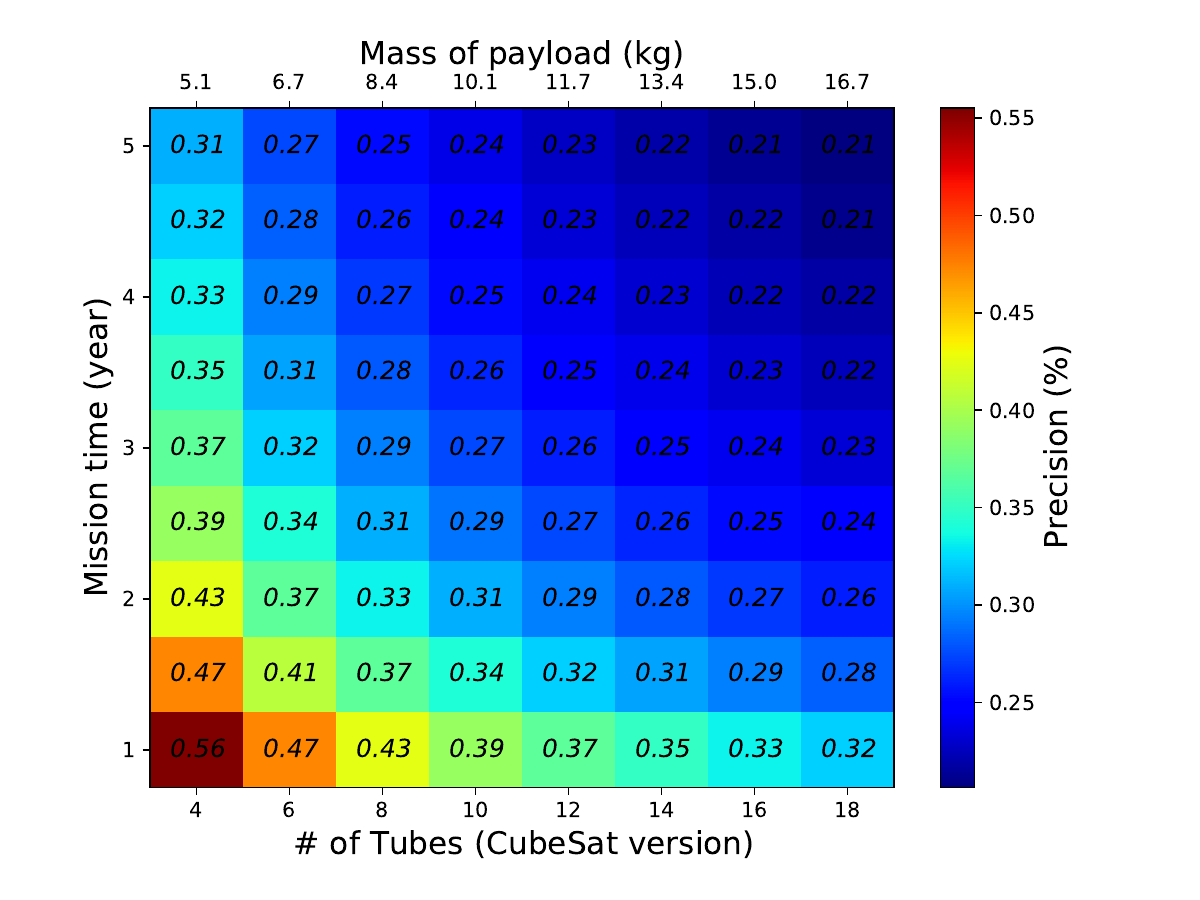}	
		\caption{Expected statistical precision of CXB measurement versus mission time and the number of tubes.}
		\label{fig:measure_precision}
	\end{figure}

Fig. \ref{fig:measure_precision} gives the resulting statistical accuracy of the CXB normalization as a function of the number of detector modules and mission time.

The CXB normalization measurement also depends on systematic uncertainties in the various spectral components considered above and on the absolute flux calibration of the instrument. As shown in Sect. \ref{sec:calibration}, the latter will be calibrated in orbit by tagged radioactive sources with an accuracy reaching 1\% every day. Accumulating calibration events over time allows to improve accuracy for longer effective exposure time. Ideally, a 100-day of exposure to the calibration source, will result in $10^6$ calibration events, i.e. a statistical accuracy of 0.1\%. While in orbit, the spectral responses (especially the absolute detection efficiency) could change because of the SAA or Solar activity, which will result in time-dependent responses. A dual ring placement of the detector tubes offers a window to study the systematics of such changes day by day and tube by tube. Changes in the response on orbital timescales, for instance following the SAA passage will be calibrated. Data with significant changes (both on the time interval and tube array) will be discarded, resulting in a loss of exposure time (or collecting area). Therefore the accuracy of absolute calibration is limited by the accuracy one can achieve on ground and the systematical variation in orbit. The very large redundancy provided by 18 tubes and the data set representing orders of magnitude more events than what the statistical error requires will allow to study and correct all the possible new systematics due to the space environment. Therefore the final uncertainty will be limited by the ground calibration performance which we estimate at 1\%.
   
Given these simulations, we expect that a CubeSat mission with 18 detector tubes operating for more than two years would allow to measure the CXB normalization with an accuracy of $\sim$1\%. 

In the 100-1000 keV band, the subtraction of the instrumental background would leave 2\% uncertainty on the CXB normalization for an 18-tube mission running for two years (based on Table \ref{tab:bkg_rates}). However, the collimators are gradually becoming  transparent above 100 keV, therefore all the known non-AGN sources and unresolved galactic components would contaminate the CXB measurement, each category introducing an uncertainty comparable to that of the instrumental background. The efficiency of detection of the 511 keV line will be very well calibrated on ground using tagged $\beta^+$ source. However, in space, each detector will develop a very large 511 keV background line due to activation. This line can be used for energy calibration but it will be very difficult to maintain an absolute efficiency calibration of every tube at high energy. Uncertainty on the CXB normalization in the 100-1000 keV range is therefore estimated at a level of $\sim$10\% 

\section{Summary and discussion}\label{sec:discussion}
	
The CXB is made of the superposition of the emission of celestial sources, mostly AGN. Numerous space missions have measured the CXB spectrum, and few of them particularly surveyed the AGN population. As a result, the CXB is nearly ($>$93\%) resolved into point-like AGN at soft X-rays (below 10 keV). A percentage decreases with increasing energy. Accurate measurement of the CXB spectrum and normalization is crucial to study the population of AGN, their obscuration, reflection, average spectra and ultimately the history of accretion in the Universe. The uncertainty on the CXB normalization ($\sim$15\%) is one of the main sources of difficulty affecting the CXB modeling. 
	
We propose a detector to determine the CXB normalization with a per cent level accuracy. The detector consists of an array of tubes with collimated spectrometers and rotating obturators modulating the signals and allowing to precisely extract the CXB photons from the background. We present here a preliminary design of the detector which could be accommodated on various platforms (16-U CubeSat, small satellite, space station).
	
The 16-U CubeSat option has been used to simulate the instrument performance with Geant4 taking into account the point sources and instrumental background to assess their respective count rates and the resulting accuracy on the CXB normalization. In two years, the CubeSat mission is able to measure it with an accuracy $\sim$1\% in the range 10-\SI{100}{keV} ultimately limited by the quality of the calibration performed before the launch. This is a significant improvement compared to the current measurements.


\newpage
\section*{Declarations}

\subsection*{Author's Contribution}
R. Walter and N. Produit initialized this project. H. Li performed the simulation and analysis. F. Hubert contributed to the design of the wheel system and CAD drawings. All authors contributed to the instrumental design, manuscript drafting, and reviewing.

\subsection*{Funding}

We acknowledge the support of the Swiss National Science Foundation. 

\subsection*{Conflicts of Interest}
The authors declare that the research was conducted in the absence of any commercial or financial relationships that could be construed as a potential conflict of interest.

\subsection*{Consent to participate} 
Not applicable.

\subsection*{Consent for publication} 
Not applicable.

\subsection*{Code availability}
Not applicable.

\bibliography{sn-bibliography.bib}
	
	

\end{document}